# Magnetoencephalography based on high-$T_c$ superconductivity: a closer look into the brain?

F. Öisjöen[*,1], J. F. Schneiderman[2,3], G. A. Figueras[1], M. L. Chukharkin[1,4], A. Kalabukhov[1,5], A. Hedström[6], M. Elam[2,3,6], D. Winkler[1].

[*]Correspondence and requests for materials should be addressed to fredrik.oisjoen@chalmers.se
[1]Chalmers University of Technology, Department of Microtechnology and Nanoscience – MC2
[2]MedTech West
[3]Sahlgrenska Academy and University of Gothenburg, Institute of Neuroscience and Physiology
[4]Moscow State University, Department of Physics
[5]Skobeltsyn Institute of Nuclear Physics, Moscow State University, Moscow, Russia
[6]Sahlgrenska University Hospital, Gothenburg, Department of Clinical Neurophysiology

**Abstract** Magnetoencephalography (MEG) enables the study of brain activity by recording the magnetic fields generated by neural currents and has become an important technique for neuroscientists in research and clinical settings [1-5]. Unlike the liquid-helium cooled low-$T_c$ superconducting quantum interference devices (SQUIDs) that have been at the heart of modern MEG systems since their invention [6], high-$T_c$ SQUIDs can operate with liquid nitrogen cooling. The relaxation of thermal insulation requirements allows for a reduction in the stand-off distance between the sensor and the room-temperature environment from a few centimeters to less than a millimeter, where MEG signal strength is significantly higher. Despite this advantage, high-$T_c$ SQUIDs have only been used for proof-of-principle MEG recordings of well-understood evoked activity [7-10]. Here we show high-$T_c$ SQUID-based MEG may be capable of providing novel information about brain activity due to the close proximity of the sensor to the head. We have performed single- and two-channel high-$T_c$ SQUID MEG recordings of spontaneous brain activity in two healthy human subjects. We demonstrate modulation of the occipital alpha rhythm and the mu rhythm found in the motor cortex. Furthermore, we have discovered uncharacteristically high-amplitude activity in the theta-band from the occipital region of the brain. Our results suggest high-$T_c$ SQUIDs can provide a closer look into the brain. We anticipate our measurements to be a starting point for more sophisticated full-head high-$T_c$ MEG investigations. A closely-spaced array of e.g. 10 sensors could be used to better localize the occipital theta-band signals and more studies are needed to understand their physiological origins. Furthermore, coherence of spontaneous brain rhythms has been identified as a marker of cognitive degeneration and Alzheimer's disease [11] and the demonstrated potential high-$T_c$ MEG has for identifying new rhythms will be relevant for such developments.

In recent years, MEG has become an increasingly important tool for researchers in basic and clinical neuroscience as the technique has become more refined. One of the most important applications of MEG is in pre-surgical localization of epileptic seizure foci, for which the technique has been used since the 1980s [12]. Other applications include mapping of somatosensory functions, estimating the integrative status of the brain, exposing mechanisms of cognitive deterioration, language lateralization and localization of speech-related activity, elucidating the principles of deep brain stimulation, and studying the activity of the cerebral cortex [1-5, 11, 13].

The first MEG recordings were made with an induction coil [14] and led to a significant leap in neuroscience research. A few years later the invention of the low-$T_c$ SQUID revolutionized the field by improving the sensitivity of magnetic recordings by orders of magnitude [6]. In modern MEG systems, hundreds of SQUID sensors are enclosed in a helmet that surrounds the subject's head and map the magnetic field emanating from the brain. Low-$T_c$ SQUIDs are preferred for state-of-the-art MEG systems because of their high fabrication yield and superior noise performance. A typical noise figure for such a SQUID is 1-5 femtoTesla/√Hz (1 fT = $10^{-15}$ T) at 10 Hz [15,

16], roughly one order of magnitude better than a similar high-$T_c$ device (see supporting information). However, in order to keep the low-$T_c$ MEG sensors operating at 4 K, thermal insulation limits the separation between the cold sensors and the room temperature environment to 18 mm at best (Elekta, Neuromag®).

The high-$T_c$ SQUID capability to operate at 77 K means the stand-off distance between the sensor and the room-temperature environment can be reduced to less than one millimeter [17, 18]. The closer proximity to the sources of brain activity then compensates for the higher noise levels of high-$T_c$ SQUID sensors. In fact, high-$T_c$ systems are theoretically capable of a higher signal-to-noise ratio (SNR) than their low-$T_c$ counterparts in magnetophysiological recordings [19]. Furthermore, the reduced stand-off distance facilitates detection of higher order magnetic sources, e.g. quadrupoles, that may be difficult to detect just a few centimeters from the scalp [20].

The first MEG recordings with high-$T_c$ SQUIDs were made in 1993 when Zhang et al. recorded the evoked response to auditory stimuli [7]. Similar studies of evoked brain activity have proven high-$T_c$ technology is sensitive enough to record signals from well-understood MEG sources [8-10]. However, novel experiments that go beyond proof-of-principle and explore new types of undocumented brain activity are necessary to indicate high-$T_c$ SQUID technology may supplement or even replace its low-$T_c$ predecessor.

In this paper, we present MEG recordings of spontaneous brain activity with single- and two-channel high-$T_c$ SQUID magnetometer systems. The SQUIDs of the two-channel system could be independently placed at nearly arbitrary locations on the head, the only limitation being that they could not be placed closer than roughly 10 cm of one another (due to the size of the cooling systems used). We demonstrate modulation of the occipital alpha rhythm via visual stimulation as well as modulation of the mu rhythm in the motor cortex via motor activation/flexion, both of which are well-understood phenomena. We have also discovered unique theta-band (4-8 Hz) activity in the occipital region of the head with unexpectedly high amplitude. Such occipital theta-band activity has, to our knowledge, not been reported with state-of-the-art MEG systems or other functional brain imaging/recording techniques (e.g. EEG).

One of the most well-known of the spontaneous brain activities is the alpha rhythm, present in the 8-13 Hz range. Alpha rhythms during wakefulness signify cortical resting activity. Occipital (i.e. from the back of the head) alpha activity is observed when visual input is blocked, e.g. when a subject is relaxed and has closed eyes, and becomes significantly attenuated upon resumption of visual input, e.g. opening of the eyes [21, 22]. In a similar manner, the mu rhythm (also 8-13 Hz) is a resting rhythm of the motor cortex and is suppressed by e.g. flexing the hands [22].

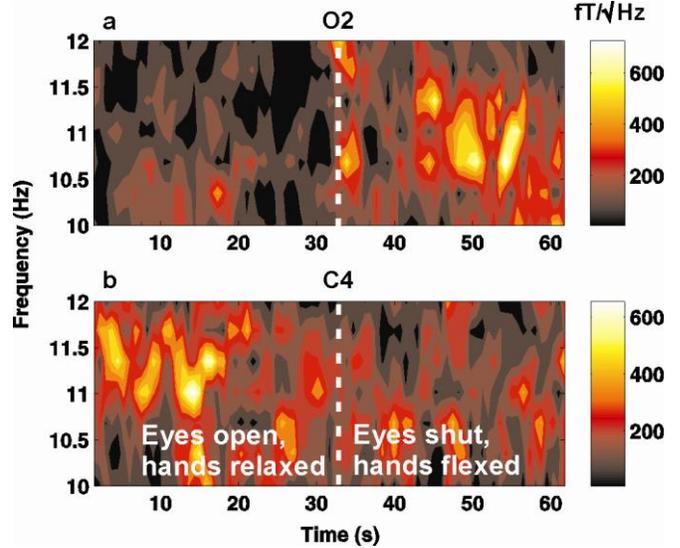

**Figure 1** Time-frequency plots of an MEG recording. The recording was made from the occipital (O2, panel **a**) and motor cortex (C4, panel **b**) of a human subject's head with two high-$T_c$ SQUID magnetometers. The subject started with open eyes and relaxed hands. After 32 seconds, the subject was instructed to flex his hands and shut his eyes. **a**, Spontaneous brain activity in the occipital region of the brain. Alpha activity at 10-12 Hz is absent during the first 32 seconds and exceeds 600 fT/√Hz upon closing the eyes, as expected. **b**, Spontaneous brain activity in the motor cortex recorded during the same 64 seconds as the alpha activity. Also as expected, the mu-rhythm is clearly attenuated when the subject flexes his hands.

Figure 1 presents a 64-second time-frequency recording obtained with our two-channel MEG system. One high-$T_c$ SQUID magnetometer was placed in the occipital region (O2/back right) of a subject's head, the other over the motor cortex (C4/right side), and both were within roughly 3 mm of the scalp surface. The subject started with open eyes and relaxed hands and was verbally instructed to both shut the eyes and flex the hands after 32 seconds. As expected, occipital alpha activity (Figure 1a) at 10-12 Hz was suppressed until the subject shut his eyes. Simultaneous recordings of mu activity from the motor cortex (Figure 1b) show strong 10-12 Hz signals that were suppressed upon flexion of the hands, also as expected. Further independent modulations of the alpha and mu rhythms were performed and corroborated these results with our MEG system and parallel EEG recordings.

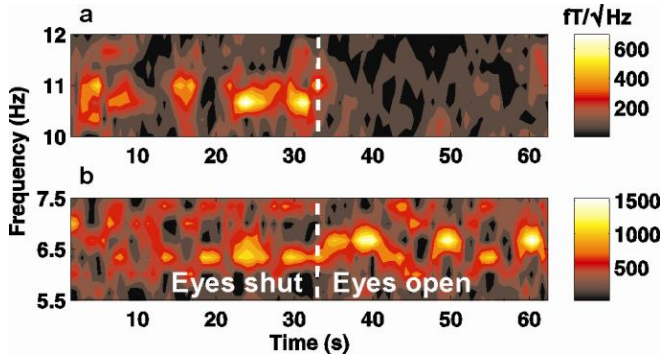

**Figure 2** Time-frequency plot in two bands of spontaneous brain activity. The plotted data was recorded from the occipital (O1) region of the head with the single-channel high-$T_c$ SQUID MEG system. The subject began with eyes shut and was verbally instructed to open them at 32 seconds. The horizontal frequency bands are plotted independently for scaling purposes. **a,** The measured alpha-band activity exceeds 600 fT/√Hz when the subject has closed eyes and is suppressed below the noise floor of the system upon opening the eyes, as expected. **b,** The theta-band signal shifts to higher frequency upon opening of the eyes and peaks with significantly high amplitude. The amplitude of the theta-band activity anomalously exceeds that of the alpha by more than a factor of 2, peaking at more than 1 pT/√Hz.

We also noted unusually high amplitude occipital activity in the theta-band (4-8 Hz). Figure 2 is an exemplary recording of such activity in which a single high-$T_c$ SQUID recorded from the occipital region (O1/back left) of the head. As expected, the alpha activity at 10-12 Hz was suppressed when the subject had open eyes (Figure 2a). Surprisingly, the amplitude of signals in the theta-band (Figure 2b) exceeds that of the alpha by roughly a factor of two whether the subject's eyes were open or shut.

Figure 3 presents a time-frequency recording with the two-channel high-$T_c$ MEG system from the occipital region (O2) and the motor cortex (C4). In this case, the subject's eyes were open throughout the recording. The mu signals in C4 (Figure 3a) were again modulated with motor activity, as expected. In this case, theta-band activity is present in both channels (Figure 3b and c) and is roughly correlated. Concurrent 32-channel electroencephalographic (EEG) recordings confirmed the mu-rhythm modulation but did not yield detectable theta-band activity from any region of the head.

The relative amplitudes of the theta-band activity (Figure 3b and c) indicate the sources for these signals were in the occipital region of the subject's head. An obvious way of understanding the nature of these occipital sources would be to map the extent of the signals they generate over the surface of the head and measure the strength of the signals as a function of distance from the head by e.g. comparing them to the strength of the alpha activity. However, the sources of the signals are not yet understood from a physiological point of view, making consistent recordings of their amplitudes unreliable.

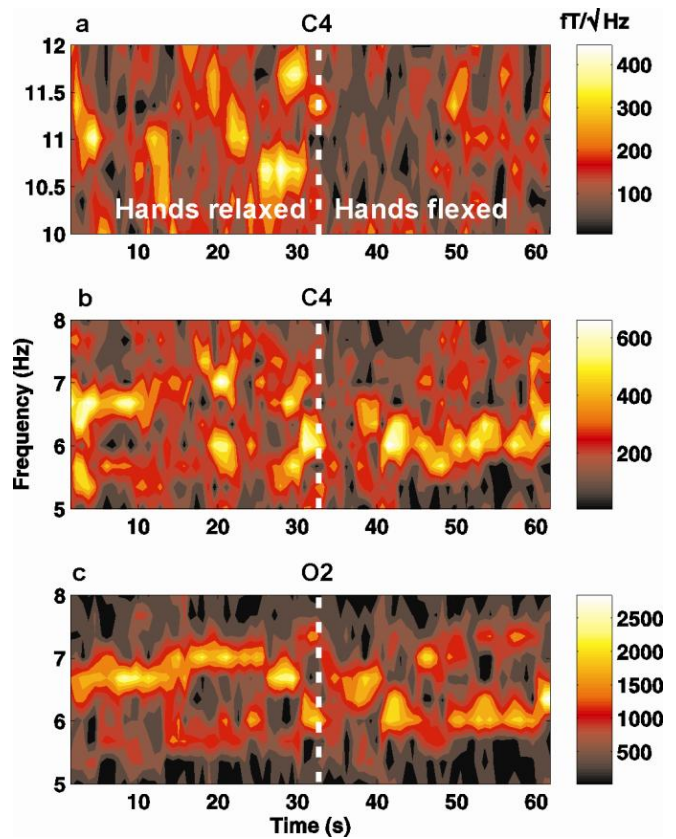

**Figure 3** Time-frequency plots of spontaneous brain activity as recorded with the two-channel high-$T_c$ SQUID MEG system from the occipital (O2) and the motor cortex (C4) regions of the brain. The subject had open eyes throughout the recording. **a,** The measured mu activity exceeds 400 fT/√Hz when the subject is relaxed (0-32 s) and is attenuated upon flexion of the hands (32-64 s), as expected. **b,** and **c,** Theta-band activity in C4 and O2, respectively. The signals are roughly correlated in the occipital and motor cortex regions of the head and two frequency peaks briefly merge when the subject is instructed to flex the hands. The amplitude of the theta-band signal in the occipital region is unexpectedly high, exceeding 2 pT/√Hz.

In general, our ability to study the nature of the theta-band activity was hindered by the fact that it was not always present, indicating the theta-band signals modulate via unknown mechanisms. However, extensive two-channel recordings in which one SQUID was located at O2 and the second SQUID recorded from other regions of the head (e.g. FP2/frontal and T4/temporal) yielded occipital theta-band activity that, on occasion, coupled weakly to other regions of the head (e.g. Figure 3b and c). When detected, the theta-band signals were consistently strongest in the occipital channel.

The alpha and mu rhythm recordings serve well as validation for our MEG capability; the detected low frequency activity in the theta-band is far more interesting from a neurophysiological point of view. Theta activity in adult subjects is normally detected at the frontal regions of the head during sleep or while performing working memory tasks [23-25] and is generally weaker than the alpha rhythm. The anomalous occipital theta-band signals, when present, were often stronger than their respective alpha counterparts, regardless of the subject's visual input or lack thereof. The occasional shifts of the theta-band activity (see Figure 2 and Figure **3**) suggest awareness or attention may modulate this rhythm. These novel occipital theta-band phenomena were reproduced in repeated recordings from two healthy adult subjects.

The common model used for MEG source localization is the equivalent current dipole model that limits brain activity to the simplest of magnetic sources [2]. By including higher order sources, more complicated patterns of mental action can be studied and understood [20]. However, the magnetic field emanating from such higher order sources decays rapidly with distance and may therefore be too weak to be detected more than a centimeter away from the surface of the head. The close proximity to the scalp we achieved with our high-$T_c$ SQUIDs may thus be responsible for our ability to record what appear to be novel sources of theta-band activity. The lack of theta-band signals in our EEG recordings also indicates these signals may arise from e.g. quadrupole sources to which EEG is blind.

The SQUID magnetometers used in our MEG recordings were made via deposition and patterning of single superconducting thin films (see supporting online material). This makes our fabrication process inherently simple compared to the multilayer structures required for better sensitivity [26]. That simplicity translates directly into high yield with our technology: we reliably fabricate SQUIDs with sufficient noise performance for incorporation in a multi-channel high-$T_c$ MEG system.

Ultimately, one should aim for a competitive advantage over the state-of-the-art in MEG systems. In modern systems, most sensors are more than 20 mm from the subject's scalp because it is rare for a person's head to fit exactly inside the rigid helmet-shaped liquid helium dewars that low-$T_c$ technology require. High-$T_c$ technology allows for both a reduction in this distance to less than a millimeter and a more flexible cooling system configuration. It is therefore possible to incorporate an array of our SQUID sensors into an MEG helmet that fits snugly around arbitrary head sizes and shapes. Such an array of sensors in close proximity to the head is likely to increase the spatial resolution of MEG recordings and enable localization of new sources of brain activity, such as the occipital theta-band signals presented herein or activity in higher frequency bands of interest [5, 27, 28]. Such a full-head high-$T_c$ SQUID-based MEG system may thus provide a closer look at the nature of brain activity.

**References:**
1. Del Gratta, C., et al., *Magnetoencephalography - a noninvasive brain imaging method with 1 ms time resolution.* Reports on Progress in Physics, 2001. **64**(12): p. 1759-1814.
2. Hamalainen, M., et al., *Magnetoencephalography - Theory, Instrumentation, and Applications to Noninvasive Studies of the Working Human Brain.* Reviews of Modern Physics, 1993. **65**(2): p. 413-497.
3. Hari, R. and O.V. Lounasmaa, *Recording and Interpretation of Cerebral Magnetic-Fields.* Science, 1989. **244**(4903): p. 432-436.
4. Chandrasekaran, C. and A.A. Ghazanfar, *When what you see is not what you hear.* Nature Neuroscience, 2011. **14**(6): p. 675-676.
5. Arnal, L.H., V. Wyart, and A.L. Giraud, *Transitions in neural oscillations reflect prediction errors generated in audiovisual speech.* Nature Neuroscience, 2011. **14**(6): p. 797-U164.

6. Cohen, D., *Magnetoencephalography - Detection of Brains Electrical-Activity with a Superconducting Magnetometer.* Science, 1972. **175**(4022): p. 664-&.
7. Y. Zhang, Y.T., M. Muck, A.I. Braginski, C. Heiden, S. Hampson, C. Pantev, T. Elbert, *Magnetoencephalography using high temperature rf SQUIDs.* Brain Topography, 1993. **5**(4): p. 4.
8. Dilorio, M.S., K.Y. Yang, and S. Yoshizumi, *Biomagnetic Measurements Using Low-Noise Integrated Squid Magnetometers Operating in Liquid-Nitrogen.* Applied Physics Letters, 1995. **67**(13): p. 1926-1928.
9. Drung, D., et al., *Integrated YBa2Cu3O7-x magnetometer for biomagnetic measurements.* Applied Physics Letters, 1996. **68**(10): p. 1421-1423.
10. Barthelmess, H.J., et al., *Low-noise biomagnetic measurements with a multichannel dc-SQUID system at 77 K.* Ieee Transactions on Applied Superconductivity, 2001. **11**(1): p. 657-660.
11. Zamrini, E., et al., *Magnetoencephalography as a Putative Biomarker for Alzheimer's Disease.* International Journal of Alzheimer's Disease. **2011**.
12. Rose, D.F., P.D. Smith, and S. Sato, *Magnetoencephalography and Epilepsy Research.* Science, 1987. **238**(4825): p. 329-335.
13. Hansen, P.C., M.L. Kringelbach, and R. Salmelin, *MEG : an introduction to methods*, New York: Oxford University Press. xii, 436 p.
14. Cohen, D., *Magnetoencephalography - Evidence of Magnetic Fields Produced by Alpha-Rhythm Currents.* Science, 1968. **161**(3843): p. 784-&.
15. Ahonen, A.I., et al., *122-Channel Squid Instrument for Investigating the Magnetic Signals from the Human Brain.* Physica Scripta, 1993. **T49a**: p. 198-205.
16. Burghoff, M., et al., *Squid System for Meg and Low Field Magnetic Resonance.* Metrology and Measurement Systems, 2009. **16**(3): p. 371-375.
17. Magnelind, P., et al., *Magnetophysiology of Brain Slices Using an HTS SQUID Magnetometer System.* Applications of Nonlinear Dynamics-Model and Design of Complex Systems, 2009: p. 323-330.
18. Öisjöen, F., et al., *A new approach for bioassays based on frequency- and time-domain measurements of magnetic nanoparticles.* Biosensors and Bioelectronics, 2010. **25**(5): p. 1008-1013.
19. Tarte, E.J., et al., *High T-c SQUID systems for magnetophysiology.* Physica C-Superconductivity and Its Applications, 2002. **368**(1-4): p. 50-54.
20. Jerbi, K., et al., *On MEG forward modelling using multipolar expansions.* Physics in Medicine and Biology, 2002. **47**(4): p. 523-555.
21. Berger, H., *Ãœber das Elektrenkephalogramm des Menschen.* European Archives of Psychiatry and Clinical Neuroscience, 1933. **98**(1): p. 231-254.
22. Hari, R. and R. Salmelin, *Human cortical oscillations: A neuromagnetic view through the skull.* Trends in Neurosciences, 1997. **20**(1): p. 44-49.
23. Inanaga, K., *Frontal midline theta rhythm and mental activity.* Psychiatry and Clinical Neurosciences, 1998. **52**(6): p. 555-566.
24. Raghavachari, S., et al., *Gating of human theta oscillations by a working memory task.* Journal of Neuroscience, 2001. **21**(9): p. 3175-3183.
25. Tesche, C.D. and J. Karhu, *Theta oscillations index human hippocampal activation during a working memory task.* Proceedings of the National Academy of Sciences of the United States of America, 2000. **97**(2): p. 919-924.
26. Faley, M.I., et al., *A New Generation of the HTS Multilayer DC-SQUID Magnetometers and Gradiometers.* 7th European Conference on Applied Superconductivity (Eucas'05), 2006. **43**: p. 1199-1202.
27. Canolty, R.T., et al., *High gamma power is phase-locked to theta oscillations in human neocortex.* Science, 2006. **313**(5793): p. 1626-1628.
28. Yizhar, O., et al., *Neocortical excitation/inhibition balance in information processing and social dysfunction.* Nature, 2011. **477**(7363): p. 171-178.

**Supplementary Information** follows.

**Acknowledgements**

The authors acknowledge financial support from the European Commission Framework Program 7 (FP7/2007–2013) project MEGMRI (grant agreement n° 200859), the European Union Regional Development Fund for MedTech West, the Knut and Alice Wallenberg foundation, and Kristina Stenborgs stiftelse. We thank Johan Wessberg, Franca Tecchio, Stefania della Penna, Vittorio Pizzella, Gian Luca Romani, and Risto Ilmoniemi for discussions.

**Author Contributions** FÖ and JFS contributed equally to the measurements and writing of this manuscript. GAF helped with the measurements. ME and AH assisted in data interpretation and experimental design. MLC helped with sensor fabrication. AK and DW assisted in writing of this manuscript and data interpretation.

# Supplementary information:

We fabricated our sensors via epitaxial growth of a thin film of $YBa_2Cu_3O_{7-x}$ (YBCO, $T_c$ = 93 K) on a strontium titanate (STO) bicrystal substrate. A single-layer 300 nm thick YBCO film was deposited by pulsed laser ablation and optical lithography was used to pattern the structures. The magnetometers had an 8×8 mm² pickup loop directly coupled to the SQUID, as shown in Supplementary Figure 1. The SQUID is a flux-to-voltage transducer and can be operated in two modes: dc bias and bias reversal [29]. Low frequency (1/$f$) noise in high-$T_c$ SQUIDs tends to be dominated by critical current fluctuations that can be cancelled with bias reversal, thus significantly improving sensitivity for MEG recordings. The magnetic field noise as a function of frequency of one of the SQUID magnetometers we used is shown in Supplementary Figure 1. At the relevant frequencies where well-documented spontaneous brain activity occurs (4-30 Hz), the noise is well below 100 fT/√Hz, coming down to just 43 fT/√Hz at 10 Hz with bias reversal. The white noise of this device was 25 fT/√Hz above 40 Hz. This noise performance was well reproducible and several devices with similar characteristics have been fabricated and tested.

We performed MEG recordings in a three-layer magnetically shielded room (MSR by Vacuum Schmeltze GmbH, with 2 mu-metal layers sandwiching a copper-coated aluminum layer) in order to reduce magnetic noise. The SQUID magnetometers were mounted on sapphire rods in a pair of non-magnetic epoxy-reinforced glass fiber cryostats (ILK Dresden). The sapphire rods were thermally connected to hermetically sealed liquid nitrogen baths that could be pumped on to controllably cool the SQUIDs between 70 and 77 K. The cold SQUIDs were separated from the room-temperature environment by 200 μm thick sapphire windows. The sensors could be manually moved towards or away from the windows. For MEG recordings, the two cryostats were mounted on a flexible dual arm setup where the cryostats could be moved to nearly arbitrary locations on the subject's head. However, the size of the cryostat heads placed a lower limit on the spacing between the individual SQUIDs of roughly 10 cm, preventing parallel recordings from the same region of the head. The SQUIDs were operated in flux-locked loop mode (Magnicon SEL-1 SQUID electronics) and their outputs were filtered and amplified (3-30 Hz or 1-30 Hz, with a gain of typically 2000) before being recorded in the time domain with a digital acquisition card (National Instruments model USB-6221).

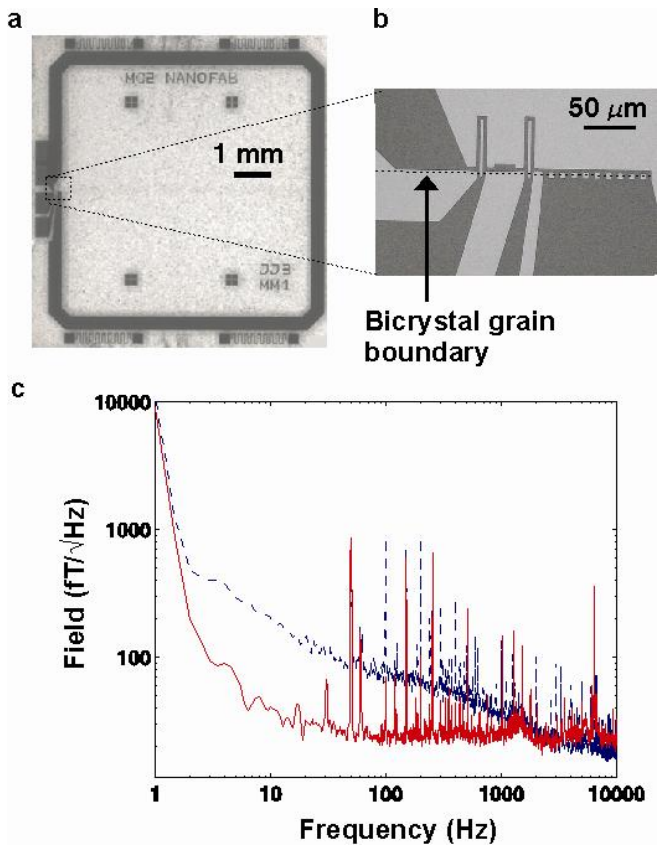

**Supplementary Figure 1** SQUID magnetometer used for MEG. **a,** Micrograph of one of the SQUID magnetometers made from YBCO. The large square is the pickup loop that feeds current into the SQUID. **b,** Zoomed-up view of the SQUIDs. Two SQUIDs are fabricated in the same pickup loop for redundancy, only one was used for each MEG channel during recordings. The bicrystal grain boundary (dashed line) crosses the 3 μm constrictions in the film, creating Josephson junctions. **c,** Equivalent magnetic field noise as a function of frequency with two biasing modes. The bias reversal mode (red, solid) significantly reduced the low frequency (1/$f$) noise as compared with the dc-bias mode (blue, dashed). The noise level of this SQUID magnetometer at 10 Hz was 43 fT/√Hz.

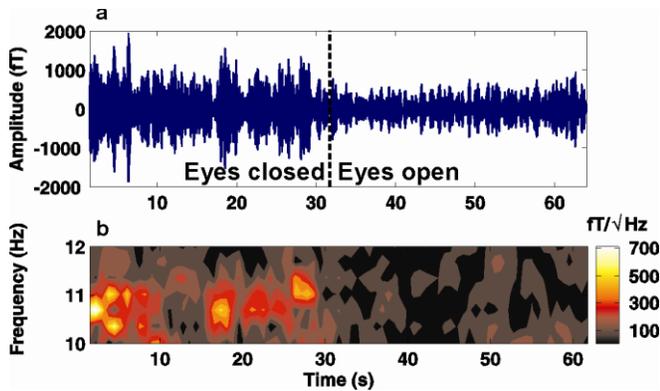

**Supplementary Figure 2** High-$T_c$ MEG recording. **a,** Time trace (band-pass filtered, 8-12 Hz) of a high-$T_c$ MEG recording from the occipital region of an awake and alert subject. Alpha activity is suppressed when the subject opens his eyes after 32 seconds. **b,** Time-frequency plot of the time-trace in panel A. Again, the alpha activity at 10.5-11 Hz is substantially attenuated after the subject opens his eyes.

Figure 1, Figure 2, & Figure 3 (main text) include analyzed data from time-domain recordings with our high-$T_c$ SQUIDs. In order to demonstrate how the analysis was made, we show a time-trace (band-pass filtered, 8-12 Hz) and the corresponding time-frequency plot of a high-$T_c$ MEG recording from the occipital region of the head in Supplementary Figure 2. The 64-second time trace (Supplementary Figure 2a) was partitioned into continuous and partially overlapping 3-second intervals with 0.7 second time steps. Each interval independently underwent a fast Fourier transform (FFT). The width of the time-domain interval (3 seconds) was selected to ensure proper density of points for the FFT and the step in the time domain (0.7 seconds) prevented undersampling artifacts. The FFT traces were concatenated into a 3D map for which the x-axis corresponds to the time-intervals, the y-axis the frequency, and the z-axis (color scale) the amplitude of the FFT (Supplementary Figure 2). The map in Figure 2 was split along the frequency axis according to the associated bands of brain activity: alpha (10-12 Hz) and theta (5.5-7.5 Hz). The splitting was necessary because of the drastically different amplitudes of the signals, i.e. had all activity been plotted in a single map, the theta-band peaks would overwhelm those associated with the alpha activity. Within the bandwidth of our measurement (1-30 Hz), no remarkable activity is present in the frequency ranges located in between or outside the plotted bands. Note that the attenuation of the alpha-rhythm can be seen in the time-trace in supplementary Figure 2a, and its frequency can be read out in the time-frequency plot (Supplementary Figure 2b).

In order to verify that the high amplitude theta activity we observed was not an artifact from the heartbeat or pulse of the subject, we recorded the theta-band activity from the occipital region of the head with the first SQUID while simultaneously recording the heartbeat with the second placed near the chest. Supplementary Figure 3 shows the frequency spectrum of such a recording. In this figure, peaks in the low-frequency theta-band arise exactly in between peaks arising from the heartbeat. The pulse was later measured on the subject's finger using an optical absorption method, also in parallel with a heartbeat recording. The pulse recording provided information about vibrations that may be induced by the pumping of blood in blood vessels close to the skull and could thus cause artifacts in the theta-band recordings. The frequency components of the pulse and the heartbeat measurements overlapped, as expected. It is thus evident that the theta-band frequency components of the occipital channel are independent of the magnetic signature of the heartbeat and are not an artifact of vibrations caused by blood flow.

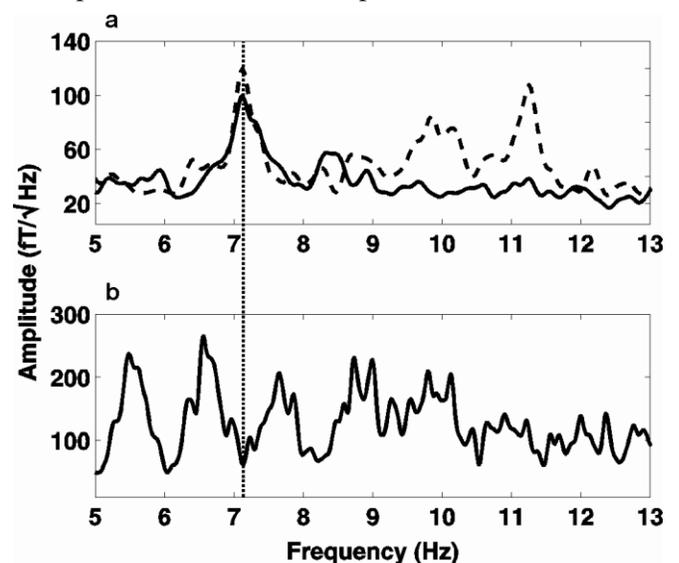

**Supplementary Figure 3** Spectra of a single 64-second MEG and MCG recording with two high-$T_c$ SQUIDs. **a,** Brain activity from the occipital region of an awake and alert subject's head. Alpha activity (9-12 Hz) is strongest when the subject's eyes are closed (dashed line) and suppressed when the subject has open eyes (solid line), as expected. The anomalous theta-band activity (peak at ~7.1 Hz) is stronger than the alpha activity and present regardless of the subject's visual input. **b,** Simultaneous SQUID-recorded heartbeat. **a,** and **b** Note the frequency components of the theta-band signal (marked with the dotted line at 7.1 Hz) and that of the heartbeat do not correspond. The theta-band activity is thus independent of the heart's magnetic signature.

**Supplementary References:**

29. Drung, D., *High-T-c and low-T-c dc SQUID electronics.* Superconductor Science & Technology, 2003. **16**(12): p. 1320-1336.